\documentclass[aip,reprint,onecolumn,jmp,graphicx]{revtex4-1}

\draft 

\usepackage{amsmath}
\usepackage{amssymb}
\usepackage{bm}

\usepackage{color}

\newcommand{\rmd}{\mathrm{d}}
\newcommand{\rme}{\mathrm{e}}

\begin{document}


\title{Karlin-McGregor-like formula in a simple time-inhomogeneous birth-death process} 



\author{Jun Ohkubo}
\email[Email address: ]{ohkubo@i.kyoto-u.ac.jp}
\affiliation{
Graduate School of Informatics, Kyoto University,\\
Yoshida Hon-machi, Sakyo-ku, Kyoto-shi, Kyoto 606-8501, Japan
}



\begin{abstract}
A novel approach is employed and developed to derive transition probabilities 
for a simple time-inhomogeneous birth-death process.
Algebraic probability theory and Lie algebraic treatments make it easy to treat the time-inhomogeneous cases.
As a result, an expression based on the Charlier polynomials is obtained,
which can be considered as an extension of a famous Karlin-KcGregor representation
for a time-homogeneous birth-death process.
\end{abstract}


\maketitle 

\section{Introduction}
\label{sec_introduction}

Birth-death processes have been widely used in various contexts including physics, biology, and social sciences 
\cite{Gardiner_book,Novozhilov2006,Aoki_book};
it is a continuous-time Markov chain
with discrete states on non-negative integers.
Not only for their applicability to modeling various phenomena,
but also for their rich mathematical structures, the birth-death processes have been studied well.
For example, the birth-death processes have discrete states,
so that a treatment based on generating functions are useful \cite{Gardiner_book};
using the generating function approach, 
various quantities, including transition probabilities, can be derived.
However, it has been shown that
orthogonal polynomials give beautiful representations
for the transition probabilities. \cite{Lederman1954,Karlin1955,Karlin1957a,Karlin1957b,Karlin1958,Schoutens_book}
The expression for the transition probabilities, so-called Karlin-McGregor spectral representation,
is based on a sequence of orthogonal polynomials and a spectral measure.
Because the orthogonal polynomials have deep relationship with continued fractions,
it would be natural to consider that the birth-death process could be dealt with by using the continued fractions.
Actually, numerical algorithms based on the continued fractions have been proposed
(for this topic, for example, see the recent paper by Crawford and Suchard. \cite{Crawford2012})

However, most of the above discussions are basically for time-homogeneous cases,
in which rate constants for the birth-death processes are time-independent.
In contrast, studies for time-inhomogeneous cases are not enough.
While the generating function approach have been applied
to time-inhomogeneous birth-death processes, \cite{Aoki_book,Kendall1948}
it has not been known even whether transition probabilities for the time-inhomogeneous birth-death processes
can be described in terms of the orthogonal polynomials or not.
The time-inhomogeneous cases are sometimes important in mathematical modeling of external influences.
In addition, in a practical sense, concise expressions for the transition probabilities
are demanded; for example, in time-series data analysis for bioinformatics, 
rapid evaluation of the transition probabilities is needed.
While we can use various Monte Carlo simulations in order to deal with the time-inhomogeneous cases,
it is important to try to find concise expressions and easy calculations for the transition probabilities.

In the present paper, we show that 
it is possible to describe transition probabilities in terms of orthogonal polynomials
at least in a simple time-inhomogeneous birth-death process.
The birth-death process has only a state-independent birth rate and linearly-state-dependent death rate.
For a time-homogeneous case of the birth-death process,
the Karlin-McGregor representation is given by the Charlier polynomials,
and our expression for the time-inhomogeneous case is also given by the Charlier polynomials.
In addition, it is possible to show that
our expression for the time-inhomogeneous case is an extension of that for the time-homogeneous case.
In order to obtain the expression in terms of the Charlier polynomials,
we employ the Lie algebraic technique proposed by Wei and Norman. \cite{Wei1963,Wei1964}
We will show that the algebraic probability theory \cite{Hora_Obata_book}
makes calculations in the Lie algebraic treatments easy and tractable.
Consequently, two different expressions for transition probabilities are obtained;
one is consistent with a result of the conventional generating function approach,
and another is the Karlin-McGregor-like formula mentioned above.

The present paper is constructed as follows.
In Sec.~\ref{sec_II}, the simple birth-death process used in the present paper
and its Karlin-McGregor representation for the time-homogeneous case are given.
Section~\ref{sec_Doi-Peliti} gives a brief review of the algebraic probability treatment 
(the so-called Doi-Peliti formulation in physics)
and a new representation for the creation and annihilation operators.
In Sec.~\ref{sec_Wei-Norman}, the Lie algebraic method developed by Wei and Norman is briefly explained. 
Section~\ref{sec_main} gives the main results of the present paper;
when we do not use any explicit representation for the creation and annihilation operators,
our theoretical treatments lead to an expression consistent with that of the generating function approach;
in contrast, when the Charlier polynomials are used as the concrete representation in the algebraic probability theory,
we finally obtain the Karlin-McGregor-like formula for transition probabilities.

\section{Model and Karlin-McGregor spectral representation}
\label{sec_II}

\subsection{A simple birth-death process (M/M/$\infty$ Queue)}
\label{sec_model}

Consider the following `reactions':
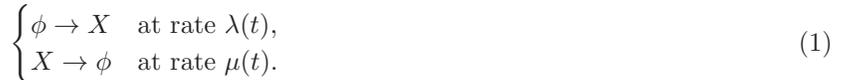
\begin{align}
\begin{cases}
\phi \to X \quad \textrm{at rate $\lambda(t)$}, \\
X \to \phi \quad \textrm{at rate $\mu(t)$}. \\
\end{cases}
\label{eq_reaction_1}
\end{align}
Such expressions of the birth-death process are sometimes used in chemical physics or population dynamics in biology.
The above `reactions' mean the following situations:
a particle $X$ is created spontaneously at rate $\lambda(t)$, and each particle $X$ is annihilated at a certain rate $\mu(t)$.
Note that since `each' particle $X$ disappears independently,
the probability of the `reaction' of the annihilation of $X$ increases with the number of particles.
Hence, one can rewritten the problem in Eq.~\eqref{eq_reaction_1} as follows: 
let $n$ be the number of the particles $X$ at time $t$, then consider the following birth-death process:
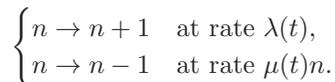
\begin{align*}
\begin{cases}
n \to n + 1 \quad \textrm{at rate $\lambda(t)$}, \\
n \to n-1 \quad \textrm{at rate $\mu(t) n$}.
\end{cases}
\end{align*}
Or, one may prefer the following definitions of the birth-death process:
Consider a Markov process with time parameter $t \in [0,\infty)$ 
on a discrete state space $S = \{0, 1, 2, \dots\}$,
in which the transition rate is defined as follows:
\begin{align*}
& \mathrm{Prob} \textrm{\{$n \to n+1$ in $(t,t+h]$\}} = \lambda(t) h + o(h) 
\quad \textrm{ as $h \downarrow 0$, for } n \in S, \\
& \mathrm{Prob} \textrm{\{$n \to n-1$ in $(t,t+h]$\}} = \mu(t) n h + o(h) 
\quad \textrm{ as $h \downarrow 0$, for } n \ge 1, \\
& \mathrm{Prob} \textrm{\{$n \to n$ in $(t,t+h]$\}} = 1 - (\lambda(t) + \mu(t) n) h + o(h) 
\quad \textrm{ as $h \downarrow 0$, for } n \in S, \\
& \mathrm{Prob} \textrm{\{$n \to m$ in $(t,t+h]$\}} = o(h) 
\quad \textrm{ as $h \downarrow 0$, for } n \in S, m \notin \{n-1, n, n+1\}.
\end{align*}

The master equation (or the Kolmogorov forward equation) is written as follows:
\begin{align}
\frac{\rmd P_{n}(t)}{\rmd t} = \lambda(t) \left[ P_{n-1}(t) - P_{n}(t)\right]
+ \mu(t) \left[ (n-1) P_{n-1}(t) - n P_n(t)\right], \quad n \in S = \{0, 1, 2, \dots\},
\label{eq_master_equation}
\end{align}
where $P_n(t)$ is the probability with $n$ populations at time $t$,
and we here used a convention of $P_{-1}(t) \equiv 0$.

The infinite-dimensional simultaneous differential equations in Eq.~\eqref{eq_master_equation} are the main problem to be solved
in the present paper.
Defining the transition probability from state $n$ at time $0$ to state $m$ at time $t$ as $P_{n \to m}(t)$,
the problem is denoted as follows:
Is it possible to represent the transition probabilities $\{P_{n \to m}(t)\}$ in a concise way,
especially in terms of a series of orthogonal polynomials?

\subsection{Karlin-McGregor spectral representation for time-homoeneous case}
\label{sec_KM_representation}

When we consider a time-homogeneous case, i.e., $\lambda(t) = \lambda$ and $\mu(t) = \mu$ for all $t$,
there is a famous representation of the transition probabilities $\{P_{n \to m}(t)\}$, 
as denoted in Sec.~\ref{sec_introduction}.
The representation, the so-called Karlin-McGregor spectral representation,
for the reactions in Eq.~\eqref{eq_reaction_1}
is given as follows \cite{Schoutens_book}:
\begin{align}
P_{n \to m}(t) = \frac{\alpha^m}{m!} \sum_{x = 0}^\infty
\rme^{- t \mu x}
C_m(x; \alpha) C_n(x; \alpha) \frac{\alpha^x}{x!} \rme^{-\alpha}
\label{eq_Karlin-McGregor},
\end{align}
where $\alpha = \lambda / \mu$
and $\{C_n(x;\alpha)\}$ is a series of the Charlier polynomials.
(For readers' convenience, a brief summary
of basic properties of the Charlier polynomials is given in the Appendix.)

\section{Algebraic representation for the birth-death process}
\label{sec_Doi-Peliti}

In order to deal with the time-inhomogeneous cases,
it is useful to employ a theoretical framework used in the algebraic probability theory. \cite{Hora_Obata_book}
More precisely, it is very convenient to introduce operators with bosonic commutation relations
in order to discuss the birth-death process.
Note that we never consider any quantum effect here;
even for the `classical' birth-death process,
the method based on the bosonic commutation relations, the so-called Doi-Peliti formulation,
has been widely used especially in statistical physics. \cite{Doi1976,Doi1976a,Peliti1985,Tauber2005}
Recently, the connection between the Doi-Peliti formulation and the algebraic probability theory has been indicated,
\cite{Ohkubo2013}
and one parameter extensions of the Doi-Peliti formulation have been proposed. \cite{Ohkubo2012}
Especially, it has been clarified that the Doi-Peliti formulation has several concrete representations,
\cite{Ohkubo2012,Droz1995}
and the relationship with the Charlier polynomials has also been suggested. \cite{Ohkubo2012}

Here, we briefly summarize the Doi-Peliti formulation.
In addition, we give a new representation based on the Charlier polynomials,
which is not given in the previous work \cite{Ohkubo2012}.

Firstly, creation operator $a^\dagger$ and annihilation operator $a$ are introduced as follows:
\begin{align}
[ a, a^\dagger] \equiv a a^\dagger - a^\dagger a = 1,
\end{align}
i.e., the creation and annihilation operators are not commute.
The action of the creation and annihilation operators on state $\{| n \rangle\}$
is defined as
\begin{align}
a^\dagger |n\rangle = |n+1\rangle, \quad a |n\rangle = n |n-1 \rangle.
\end{align}

Instead of the probability of the state $n$ in Eq.~\eqref{eq_master_equation}, $P_n(t)$,
the following ket state $| P(t) \rangle$ is used in the Doi-Peliti formalism:
\begin{align}
|P(t)\rangle \equiv \sum_{n=0}^\infty P_n(t) | n \rangle.
\label{eq_definition_of_state}
\end{align}
Using this `summarized' state $| P(t) \rangle$, various calculations become simpler and easier;
we will see them in Sec.~\ref{sec_main}.

For the Doi-Peliti formulation,
it is necessary to define suitable `bra' states (dual states for $|n\rangle$).
According to the previous work for the one-parameter extension \cite{Ohkubo2012},
we here define the following action of the creation and annihilation operators on `bra' states:
\begin{align}
\langle n| a^\dagger = \langle n-1 | n \alpha^{-1}, \quad
\langle n| a = \langle n+1 | \alpha.
\label{eq_definition_of_bra_states}
\end{align}
Hence, the orthogonality between the `bra' and `ket' states becomes as follows:
\begin{align}
\langle m | n \rangle = \alpha^{-n} n! \delta_{m,n}.
\end{align}
If $\alpha = 1$, the conventional Doi-Peliti formulation is recovered.

The above formulation is a kind of abstract one;
actually, the Doi-Peliti formulation is usually used 
without specifying explicit representations for the state $|n\rangle$,
the creation operator $a^\dagger$, and the annihilation operator $a$.
However, it is possible to obtain explicit representations for the formulation. \cite{Ohkubo2012}
There are several representations,
such as, a representation based on the correspondence with the generating function approach,
that based on the Hermite polynomials or the Charlier polynomials.
Since we define the `bra' states as Eq.~\eqref{eq_definition_of_bra_states},
the following explicit representation is obtained, which has not been proposed in the previous work 
in Ref.~\onlinecite{Ohkubo2012}.
(A different definition of the Charlier polynomials has been used in Ref.~\onlinecite{Ohkubo2012}.)
That is, for the `bra' and `ket' states,
\begin{align}
| n \rangle \equiv C_n(x;\alpha), \quad
\langle m | \equiv \sum_{x=0}^\infty \frac{\alpha^x}{x!} \rme^{-\alpha} C_m(x;\alpha),
\label{eq_concrete_representation_states}
\end{align}
and actions of the creation and annihilation operators are defined as
\begin{align}
a^\dagger f(x)= f(x) - \frac{x}{\alpha} f(x-1), \quad
a f(x) = \alpha f(x) - \alpha f(x+1),
\label{eq_concrete_representation_operators}
\end{align}
respectively.
Actually, using the basic properties of the Charlier polynomials (see the Appendix),
for example, we have
\begin{align}
a^\dagger |n \rangle = C_n(x;\alpha) - \frac{x}{\alpha}C_n(x-1;\alpha) = C_{n+1}(x;\alpha).
\end{align}

\section{Brief review of Lie algebraic method for time-inhomogenous cases}
\label{sec_Wei-Norman}

In this section, we will shortly explain
the Lie algebraic method developed by Wei and Norman. \cite{Wei1963,Wei1964}
The method by Wei and Norman has been used in various contexts,
including the Fokker-Planck equation \cite{Wolf1988} and financial topics \cite{Lo2001}.
The method has recently been applied even to the birth-death process. \cite{House2012}
However, in Ref.~\onlinecite{House2012},
an infinite-dimensional matrix have been used;
on the contrary, we use the creation and annihilation operators in the present paper,
and these notations have a kind of flexibility and availability and enable us to obtain 
the Karlin-McGregor-like formula, as shown later.

Let $\mathcal{L}$ be the Lie algebra generated by $H_1, \dots, H_L$ under the commutator product.
We assume that $\mathcal{L}$ is of finite dimension $L$.
For later use, we define an adjoint operator, $\mathrm{ad}$,
which is a linear operator on $\mathcal{L}$ and
\begin{align}
&(\mathrm{ad} H_i) H_j \equiv [H_i, H_j] = H_i H_j - H_j H_i,\\
&(\mathrm{ad} H_i)^2 H_j = [H_i, [H_i, H_j] ],
\end{align}
and so on.

We assume that the time-evolution equation for the state $| P(t) \rangle$,
defined by Eq.~\eqref{eq_definition_of_state},
is given as follows:
\begin{align}
\frac{\rmd }{\rmd t} |P(t)\rangle = H(t) |P(t) \rangle.
\label{eq_time_evolution_of_the_state}
\end{align}
As shown in Sec.~\ref{sec_main_basics}, it is possible to obtain the operator $H(t)$
in terms of the creation and annihilation operators.

Instead of the state $| P(t) \rangle$,
we here consider the time-evolution operator $U(t)$, which satisfies
\begin{align}
\frac{\rmd }{\rmd t} U(t)  = H(t) U(t)
\label{eq_time_evolution_equation_for_U_1}
\end{align}
and $U(0) = I$. ($I$ is the identity operator.)
Using the time-evolution operator $U(t)$,
the state $|P(t) \rangle$ is given as
\begin{align}
|P(t) \rangle = U(t) | P(0) \rangle.
\end{align}

The Wei-Norman method is applicable when the operator $H(t)$ can be written as
\begin{align}
H(t) = \sum_{k=1}^K a_k(t) H_k,
\end{align}
where $K$ is finite and $K \le L$. 
(Note that the Lie algebra $\mathcal{L}$ must also have a finite-dimension $L$ for the Wei-Norman method.)

Our aim here is to find an expression of the time-evolution operator $U(t)$
of the following form:
\begin{align}
U(t) = \exp\left( g_1(t) H_1 \right) \exp\left( g_2(t) H_2 \right) \cdots \exp\left( g_L(t) H_L \right),
\label{eq_form_U}
\end{align}
where $g_l(0) = 0$ for all $l \in \{1,2,\dots, L\}$.
The time derivative of Eq.~\eqref{eq_form_U} gives
\begin{align}
\frac{\rmd }{\rmd t} U(t)  = \sum_{l=0}^L \dot{g}_l(t) 
\left( \prod_{j=1}^{l-1} \exp(g_j(t) H_j) \right) H_i \left( \prod_{j=i}^{L} \exp(g_j(t) H_j) \right).
\label{eq_time_evolution_equation_for_U_2}
\end{align}
Performing a post-multiplication by the inverse operator $U^{-1}$, 
and employing the Baker-Campbell-Hausdorff formula,
\begin{align}
\rme^{H_i} H_j \rme^{- H_i} = \rme^{(\mathrm{ad} H_i)} H_j,
\end{align}
we obtain
\begin{align}
\left( \frac{\rmd }{\rmd t} U(t) \right) U^{-1}(t)
= \sum_{l=0}^L \dot{g}_l(t) 
\left( \prod_{j=1}^{l-1} \exp\left( g_j(t) (\mathrm{ad} H_j) \right) \right) H_l.
\label{eq_time_evolution_equation_for_U_3}
\end{align}
On the other hand,
we here focus on the fact that the time-evolution equation in Eq.~\eqref{eq_time_evolution_equation_for_U_1}
can be rewritten as
\begin{align}
\frac{\rmd }{\rmd t} U(t)  = \sum_{l=0}^L a_l(t) H_l U(t),
\label{eq_time_evolution_equation_for_U_4}
\end{align}
where $a_l(t) \equiv 0$ for $l > K$.
Hence, the post-multiplication by the inverse operator $U^{-1}$ gives
\begin{align}
\left( \frac{\rmd }{\rmd t} U(t) \right)U^{-1}(t)  = \sum_{l=0}^L a_l(t) H_l,
\label{eq_time_evolution_equation_for_U_5}
\end{align}
and as a result, 
the following identity is obtained by comparing
Eqs.~\eqref{eq_time_evolution_equation_for_U_3} with \eqref{eq_time_evolution_equation_for_U_5}:
\begin{align}
\sum_{l = 0}^L a_l(t) H_l = \sum_{l=0}^L \dot{g}_l(t) 
\left( \prod_{j=1}^{l-1} \exp\left( g_j(t) (\mathrm{ad} H_j) \right) \right) H_l.
\label{eq_result_wei_norman_1}
\end{align}
That is, we have a linear relation between $a_l(t)$ and $\dot{g}_l(t)$.
For more rigorous discussions, see the original papers by Wei and Norman. \cite{Wei1963, Wei1964}

The important point here is as follows:
Initially we have infinite-dimensional simultaneous coupled differential equations,
but the Wei-Norman method gives
only a finite-dimensional (at most $L$) simultaneous coupled differential equations.

\section{Two Different Expressions for Transition Probabilities}
\label{sec_main}

\subsection{The birth-death process in terms of Lie algebra}
\label{sec_main_basics}

According to the method explained in Sec.~\ref{sec_Wei-Norman},
we perform the following calculations for the model in Sec.~\ref{sec_model}.
Firstly, the operator $H(t)$ in the time-evolution equation for the state $| P(t) \rangle$
is written as follows
\begin{align}
H(t) = \lambda(t) a^\dagger - \lambda(t) I + \mu(t) a - \mu a^\dagger a.
\end{align}
Note that it is easy to check that this definition recovers the original master equation 
in Eq.~\eqref{eq_master_equation} adequately. 
(Multiply $\langle n |$ from the left side in Eq.~\eqref{eq_time_evolution_of_the_state}.)

Secondly, assuming that the state $| P(t) \rangle$ can be written as
\begin{align}
|P(t)\rangle = \rme^{g_1(t) I} \rme^{g_2(t) a^\dagger} \rme^{g_3(t) a} \rme^{g_4(t)a^\dagger a} |P(0)\rangle,
\end{align}
the following four equations are obtained from the Wei-Norman method:
\begin{align}
\begin{cases}
H_1 = I & : \quad -\lambda(t) = \dot{g}_1(t) - g_2(t) \dot{g}_3(t) - g_2(t) g_3(t) \dot{g}_4(t), \\
H_2 = a^\dagger & : \quad  \lambda(t) = \dot{g}_2(t) - g_2(t) \dot{g}_4(t),\\
H_3 = a &: \quad  \mu(t) = \dot{g}_3(t) + g_3(t) \dot{g}_4(t),\\
H_4 = a^\dagger a &: \quad  -\mu(t) = \dot{g}_4(t).
\end{cases}
\end{align}
In addition, after some calculations, it will be clarified that $g_1(t) = - g_2(t)$.

Finally, we will calculate the transition probabilities.
Let $n$ be the initial state;
i.e., set $|P(0)\rangle = |n\rangle$.
Then, the transition probability from state $n$ at time $0$
to state $m$ at time $t$ is calculated by
\begin{align}
P_{n \to m}(t) = \frac{\alpha^m}{m!} \langle m | U(t) | n \rangle.
\label{eq_transition_prob}
\end{align}

In the following subsections, we will give explicit expressions
for Eq.~\eqref{eq_transition_prob}.

\subsection{Expression 1: Finite summation expression based on abstract discussions}
\label{sec_main_expression_1}

In the Doi-Peliti formulation explained in Sec.~\ref{sec_Doi-Peliti},
the parameter $\alpha$ can be chosen arbitrarily.
For simplicity, we here assume $\alpha = 1$,
i.e., the conventional Doi-Peliti formulation.
In the following calculations, we consider the left-actions of the operators on $\langle m |$.

Firstly, we have the following expressions
up to the second factor of the time-evolution operator $U(t)$:
\begin{align}
\frac{1}{m!} \langle m | \rme^{g_1(t) I} \rme^{g_2(t) a^\dagger}
&= \rme^{g_1(t)} \frac{1}{m!} \langle m | \sum_{i=0}^\infty \frac{1}{i!} (g_2(t))^i (a^\dagger)^i
= \rme^{g_1(t)} \sum_{i=0}^\infty \frac{1}{(m-i)!} \frac{1}{i!} (g_2(t))^i \langle m-i |.
\end{align}
Secondly,
\begin{align}
\frac{1}{m!} \langle m |  \rme^{g_1(t) I} \rme^{g_2(t) a^\dagger} \rme^{g_3(t) a}
&= \rme^{g_1(t)} \sum_{i=0}^\infty \frac{1}{(m-i)!} \frac{1}{i!} (g_2(t))^i \langle m-i |
 \sum_{j=0}^\infty \frac{1}{j!} (g_3(t))^j a^j \nonumber \\
&= \rme^{g_1(t)} \sum_{i=0}^\infty  \sum_{j=0}^\infty  \frac{1}{(m-i)!} \frac{1}{i!} \frac{1}{j!} (g_2(t))^i (g_3(t))^j \langle m-i+j |.
\end{align}
Hence,
\begin{align}
&\frac{1}{m!} \langle m |  \rme^{g_1(t) I} \rme^{g_2(t) a^\dagger} \rme^{g_3(t) a} \rme^{g_4(t) a^\dagger a} |n\rangle \nonumber \\
&= \rme^{g_1(t)} \rme^{g_4(t) n}
\sum_{i=0}^\infty  \sum_{j=0}^\infty  \frac{1}{(m-i)!} \frac{1}{i!} \frac{1}{j!} (g_2(t))^i (g_3(t))^j \langle m-i+j | n \rangle \nonumber \\
&= \rme^{g_1(t)} \rme^{g_4(t) n}
\sum_{k=0}^m  \sum_{j=0}^\infty  \frac{1}{k!} \frac{1}{(m-k)!} \frac{1}{j!} (g_2(t))^{m-k} (g_3(t))^j \langle k+j | n \rangle 
\qquad \textrm{(replaced with $k = m-i$)} \nonumber \\
&= \rme^{g_1(t)} \rme^{g_4(t) n}
\sum_{k=0}^m  \sum_{l=0}^n  \frac{1}{k!} \frac{1}{(m-k)!} \frac{1}{(n-l)!} (g_2(t))^{m-k} (g_3(t))^{n-l} \langle k+n-l | n \rangle 
\qquad \textrm{(replaced with $l = k+j$)}.
\end{align}
Finally, noting the fact that $\langle k+n-l | n \rangle \neq 0$ means $k-l = 0$,
the following expression for the transition probabilities is derived:
\begin{align}
P_{n\to m}(t)
= 
\rme^{g_1(t)} \rme^{g_4(t) n}
\sum_{l=0}^{\min (m,n)}  \frac{n!}{l! (m-l)! (n-l)!} (g_2(t))^{m-k} (g_3(t))^{n-l} .
\label{eq_result_generating_function}
\end{align}

We give some comments.
In Ref.~\onlinecite{House2012}, a similar discussion has been performed
using an infinite-dimensional matrix (not the creation and annihilation operators as in the present paper),
and the transition probability for the simplest case of the initial state ($n = 0$) is derived.
In addition, as pointed out in Ref.~\onlinecite{House2012},
it would be possible to obtain the same result by using the generating function approach.
However, as shown above, the algebraic probability theory (the Doi-Peliti formulation)
gives an easy calculation way for general initial conditions.
In addition, when we consider a time-homogeneous case,
it is possible to confirm that
Eq.~\eqref{eq_result_generating_function}
reduces to the transition probabilities for the time-homogeneous case
obtained by the generating function approach (Eq.(11.1.38) in Ref.~\onlinecite{Gardiner_book}).

\subsection{Expression 2: Karlin-McGregor-like formula based on Charlier polynomials}
\label{sec_main_expression_2}

While we assume $\alpha = 1$ in Sec.~\ref{sec_main_expression_1}, 
here we consider a general case, i.e., $\alpha$ can take a certain real value.
In addition, in Sec.~\ref{sec_main_expression_1}, 
we did not use any explicit representation for the Doi-Peliti formulation;
in contrast, we here use the explicit representation based on the Charlier polynomials,
given in Sec.~\ref{sec_Doi-Peliti}.

Firstly, we act $\rme^{g_1(t) I}$ in $U(t)$ to $\langle m |$
and $\rme^{g_4(t) a^\dagger a}$ in $U(t)$ to $| n \rangle$.
Then,
\begin{align}
P_{n\to m}(t) 
&=\frac{\alpha^m}{m!} \langle m |  \rme^{g_1(t) I} \rme^{g_2(t) a^\dagger} \rme^{g_3(t) a} \rme^{g_4(t) a^\dagger a} |n\rangle \nonumber \\
&= \rme^{g_1(t)} \rme^{g_4(t) n}
\frac{\alpha^m}{m!} \langle m |  \rme^{g_2(t) a^\dagger} \rme^{g_3(t) a}  |n\rangle \nonumber \\
&= \rme^{g_1(t)} \rme^{g_4(t) n}
\frac{\alpha^m}{m!} \langle m |  \rme^{-g_2(t) g_3(t)} \rme^{g_3(t) a} \rme^{g_2(t) a^\dagger}   |n\rangle,
\end{align}
where we commuted $\rme^{g_3(t) a}$ and $\rme^{g_2(t) a^\dagger}$,
and then $\rme^{-g_2(t) g_3(t)} $ emerged:
\begin{align}
\rme^{g_2(t) a^\dagger} \rme^{g_3(t) a}
&= \exp\left( g_3(t) a + [g_2(t) a^\dagger, g_3(t) a] + \frac{1}{2!} \left[g_2(t) a^\dagger, [g_2(t) a^\dagger, g_3(t) a] \right] 
+ \cdots \right) 
\exp\left( g_2(t) a^\dagger \right) \nonumber \\
&= \exp\left( g_3(t) a - g_2(t) g_3(t) \right) \exp\left( g_2(t) a^\dagger \right) .
\end{align}

Secondly, the following convention, i.e., the coherent state, is introduced:
\begin{align}
| z \rangle \equiv \rme^{z a^\dagger} | 0 \rangle.
\end{align}
The coherent state plays an important role when we construct a path-integrals (for example, see Ref.~\onlinecite{Tauber2005}),
and the coherent state is also characterized by the following fact:
\begin{align}
a | z \rangle = z | z\rangle,
\end{align}
i.e., the coherent state is an eigen state of the annihilation operator $a$.
Hence, we obtain
\begin{align}
P_{n\to m}(t) 
&= \rme^{g_1(t)} \rme^{g_4(t) n} 
\frac{\alpha^m}{m!} \langle m |  \rme^{-g_2(t) g_3(t)}  \rme^{g_3(t) a} \rme^{g_2(t) a^\dagger}  (a^\dagger)^n |0\rangle \nonumber \\
&= \rme^{g_1(t)} \rme^{g_4(t) n} 
\frac{\alpha^m}{m!} \langle m |  \rme^{-g_2(t) g_3(t)}  \rme^{g_3(t) a} (a^\dagger)^n | g_2(t) \rangle.
\end{align}
where $| g_2(t) \rangle$ is the coherent state $| z \rangle$ with $z = g_2(t)$.

It is easy to verify the following useful identities by mathematical induction:
\begin{align}
\rme^{z a} (a^\dagger)^n = (z+a^\dagger)^n e^{z a}
\end{align}
and then we have
\begin{align}
&\rme^{-g_2(t) g_3(t)}  \rme^{g_3(t) a} (a^\dagger)^n | g_2(t) \rangle \nonumber \\
&= \rme^{-g_2(t) g_3(t)} (g_3(t)+a^\dagger)^n \rme^{g_3(t) a} | g_2(t) \rangle \nonumber \\
&= \rme^{-g_2(t) g_3(t)} (g_3(t)+a^\dagger)^n \rme^{g_3(t) g_2(t)} | g_2(t) \rangle \nonumber \\
& = (g_3(t)+a^\dagger)^n  | g_2(t) \rangle
\end{align}

Up to now, we did not use the explicit representation of the Doi-Peliti formulation.
In the following discussions, the representations given in 
Eqs.~\eqref{eq_concrete_representation_states} and \eqref{eq_concrete_representation_operators}
are necessary.
Note that the coherent state is equal to the generating function of the Charlier polynomials,
i.e.,
\begin{align}
|g_2(t)\rangle = \sum_{n=0}^\infty C_n(x;\alpha) \frac{(g_2(t))^n}{n!}= \rme^{g_2(t)} \left( 1 - \frac{g_2(t)}{\alpha} \right)^x.
\end{align}
Next, we consider an action of a operator $g_3(t)+a^\dagger$
on $| g_2(t) \rangle C_n(x;\alpha)$.
Here, the creation operator should be interpreted as in Eq.~\eqref{eq_concrete_representation_operators},
and therefore
\begin{align}
(g_3(t)+a^\dagger)  | g_2(t) \rangle C_n(x;\alpha)
&= (g_3(t)+1) \rme^{g_2(t)} \left( 1 - \frac{g_2(t)}{\alpha} \right)^x C_n(x;\alpha)
- \frac{x}{\alpha} \rme^{g_2(t)} \left( 1 - \frac{g_2(t)}{\alpha} \right)^{x-1} C_n(x-1;\alpha) \nonumber \\
&= \rme^{g_2(t)} \left( 1 - \frac{g_2(t)}{\alpha} \right)^{x-1} 
\left\{
(g_3(t)+1) \left( 1 - \frac{g_2(t)}{\alpha} \right) C_n(x;\alpha) - \frac{x}{\alpha} C_n(x-1;\alpha)
\right\}.
\end{align}
Note that if $(g_3(t)+1) \left( 1 - \frac{g_2(t)}{\alpha} \right) = 1$,
we obtain the following simple identity:
\begin{align}
(g_3(t)+a^\dagger)  | g_2(t) \rangle C_n(x;\alpha)
&=  \left( 1 - \frac{g_2(t)}{\alpha} \right)^{-1} | g_2(t) \rangle C_{n+1}(x;\alpha).
\label{eq_key_identity}
\end{align}
That is, since $\alpha$ can be chosen arbitrarily, we here set $\alpha$ as follows.
\begin{align}
\alpha = \frac{g_2(t) (g_3(t) +1)}{g_3(t)}
\end{align}
Using the fact that $C_0(x;\alpha) = 1$,
and employing Eq.~\eqref{eq_key_identity} repeatedly,
we have
\begin{align}
(g_3(t)+a^\dagger)^n  | g_2(t) \rangle 
= \rme^{g_2(t)}  \left( g_3(t) + 1 \right)^{-x+n} C_n(x;\alpha).
\end{align}
Hence, we finally obtain
\begin{align}
P_{n\to m}(t) =  \frac{\alpha^m}{m!}
\sum_{x=0}^\infty \rme^{g_4(t) n} 
C_m(x;\alpha)  C_n(x;\alpha)
\frac{\alpha^x}{x!} \rme^{-\alpha} \left( g_3(t) + 1 \right)^{-x+n}.
\label{eq_main_result}
\end{align}
where we used the fact that $g_1(t) = - g_2(t)$.

The final expression, i.e., Eq.~\eqref{eq_main_result}, is the main result of the present paper.
This is expressed in terms of the Charlier polynomials,
and it has a very similar form with the Karlin-McGregor representation for the time-homogeneous case.
In fact, when we consider the time-homogeneous case, we have
\begin{align}
g_3(t) = \rme^{\mu t} - 1, \quad g_4(t) = - \mu t,
\end{align}
which gives the same consequence with Eq.~\eqref{eq_Karlin-McGregor}.
Hence, Eq.~\eqref{eq_main_result} can be considered
as an extension of the time-homogeneous case.

\section{Concluding remarks}
\label{sec_conclusions}

In the present paper, 
a theory for the time-inhomogeneous birth-death processes was developed,
and the two different expressions for the transition probabilities were derived
for a simple birth-death process.
As shown in the present paper, the usage of the algebraic probability theory (the Doi-Peliti formulation)
gives convenient calculation methods,
and we can obtain the Karlin-McGregor-like formula
even for the time-inhomogeneous case.

The present paper showed that even in the time-inhomogeneous case,
there is at least one birth-death process in which the Karlin-McGregor-like formula exists.
Then, the following natural question arises:
Is it possible to derive similar expressions based on orthogonal polynomials
even for various birth-death processes?
This issue is under investigation.
When we can construct an adequate Lie algebra, creation and annihilation operators,
it is expected that the discussion given in the present paper is applicable for
other birth-death processes.
Especially, the usage of the coherent state would play an important role in the calculations.


\section*{ACKNOWLEDGMENTS}

This work was supported in part by grant-in-aid for scientific research 
(No.~25870339)
from the Ministry of Education, Culture, Sports, Science and Technology (MEXT), Japan.

\appendix
\section{Some basic properties of the Charlier polynomials}
\label{app_section_charlier}

The Charlier polynomials satisfy the following some properties 
\cite{Schoutens_book,Koekoek_book}.
\begin{itemize}
\item Orthogonality relation:
\begin{align}
\sum_{x=0}^\infty \frac{\alpha^x}{x!} \rme^{-\alpha} C_m(x;\alpha) C_n(x;\alpha) = \alpha^{-n} n! \delta_{mn}.
\end{align}
\item Recurrence relation:
\begin{align}
-x C_n(x;\alpha) = \alpha C_{n+1}(x;\alpha) - (n+1) C_{n}(x;\alpha) + n C_{n-1}(x;\alpha).
\end{align}
\item Generating function:
\begin{align}
\sum_{n = 0}^\infty C_n(x; \alpha) \frac{z^n}{n!} = \rme^z \left(1 - \frac{z}{\alpha} \right)^{x}.
\end{align}
\item Forward shift:
\begin{align}
C_n(x+1;\alpha) = C_n(x;\alpha) = - \frac{n}{\alpha} C_{n-1}(x;\alpha).
\end{align}
\item Backward shift:
\begin{align}
C_n(x;\alpha) - \frac{x}{\alpha} C_n(x-1;\alpha) = C_{n+1;\alpha}.
\end{align}
\item Duality relation:
\begin{align}
C_n(x;\alpha) = C_x(n;\alpha).
\end{align}
\end{itemize}

\end{document}